\documentclass{sig-alternate}

\DeclareRobustCommand{\ttfamily}{\fontencoding{T1}\fontfamily{lmtt}\selectfont}
\usepackage{ifthen}
\usepackage{amssymb}
\usepackage{footmisc}

\makeatletter
\def\@copyrightspace{\relax}
\makeatother

\begin{document}

\title{Three Metrics to Explore the Openness of GitHub projects
  \titlenote{This work was submitted to the 11$^{th}$ Working Conference on Mining Software Repositories (MSR'14), Mining Challenge track. The paper was eventually rejected. This work is licensed under CC BY 3.0.}}

\numberofauthors{1}
\author{
\alignauthor
Valerio Cosentino, Javier Luis C\'anovas Izquierdo and Jordi Cabot\\
       \affaddr{AtlanMod, EMN -- Inria -- Lina, Nantes, France }\\
       \email{\{valerio.cosentino,javier.canovas,jordi.cabot\}@inria.fr}
}
\date{17 January 2014}

\maketitle
\begin{abstract}
Open source software projects evolve thanks to a group of volunteers that help in their development. Thus, the success of these projects depends on their ability to attract (and keep) developers. We believe the openness of a project, i.e., how easy is for a new user to actively contribute to it, can help to make a project more attractive. To explore the openness of a software project, we propose three metrics focused on: (1) the distribution of the project community, (2) the rate of acceptance of external contributions and (3) the time it takes to become an official collaborator of the project. We have adapted and applied these metrics to a subset of GitHub projects, thus giving some practical findings on their openness.
\end{abstract}

\category{H.4.0}{Information Systems Applications}{General}
\category{D.2.8}{Software Engineering}{Metrics}
\category{J.4}{Computer Applications}{Social and Behavioral Sciences}

\terms{Measurement}

\keywords{Openness, Open Source Software, GitHub}

\section{Introduction}
\label{sec:introduction}
One of the main advantages of open source software is to enable anyone to reuse and modify the code without depending on the original developers \cite{Bein2010}. Open source projects provide different ways to contribute (e.g., proposing new features, applying code changes, etc.), where some of them may require the approval of the project members to be actually merged into the original project. In this sense, open source projects can be considered more or less \emph{open} according to the type of contributions they allow from non-project members. 

Openness is generally defined as \emph{a tendency to accept new ideas, methods or changes}\footnote{\texttt{http://www.macmillandictionary.com/dictionary/brit} \texttt{ish/openness}}. This definition can be adapted to the field of software development as the tendency to accept changes from non-project members and to allow them to participate in the decision-making process of the project. The development of open source projects capitalizes on developers contributing voluntarily. Thus, the vitality and success of these projects depend on their ability to attract, absorb and retain new developers \cite{Bird2007}. We believe that the openness level of an open source project can influence the developers' willingness to contribute to a project.

There are very few works trying to come up with metrics to evaluate the openness of projects. In this paper we propose three new metrics to allow gaining more insights into the openness level, specifically:

\begin{description}
  \item[M1] Community composition. \emph{How the community of the project is composed in terms of project and non-project members?}
  \item[M2] External contribution analysis. \emph{How many external contributions are accepted? How long does it take to evaluate an external contribution?}
  \item[M3] Time to become collaborator. \emph{How long does it take to become collaborator?}
\end{description}

In this paper we adapt and apply the previous metrics to a subset of projects in GitHub, that is one of the largest and best-known social coding sites. We believe our results could then be integrated as part of the project information in GitHub and also part of the information provided in other websites such as Ohloh\footnote{\texttt{http://www.ohloh.net}} or Openhatch\footnote{\texttt{http://openhatch.org}}, which provide ranking mechanisms (e.g., most popular projects, most active projects, etc.) to aid developers to look for an open source project to contribute to. 

The paper is structured as follows. Section \ref{sec:dataset} describes briefly the dataset. Sections \ref{sec:op1}, \ref{sec:op2} and \ref{sec:op3} introduce the three metrics proposed and present the results obtained from the dataset. Section \ref{sec:relatedWork} discusses the related work and finally Section \ref{sec:conclusion} ends the paper and presents some future work.

\section{Data and Approach}
\label{sec:dataset}
This year the MSR Challenge focuses on a subset of GitHub projects. GitHub is a web-based hosting service for software development projects using the Git control system. The service also supports both issue tracking capabilities and social features (e.g., followers and watchers). This section briefly describes the main concepts of the dataset to be used in our analysis later on.

Projects and users can be considered the main assets of the dataset. A project is managed by two kinds of users: the owner and the project members, the former has full control over the repository, whereas the latter are granted full management permissions. Each project keeps track of the submitted issues, pull requests (together with the corresponding associated events, e.g, creation, merging, etc.) and commits. Any user can create issues and pull requests to point out bugs or request/contribute new features. Pull requests are the main means to contribute to a project. To create a pull request, the user has first to fork a project, then perform some changes and finally send the pull request to the original project. The pull request is analyzed (some discussions can arise as comments in the pull request) and if eventually accepted, the related code changes (i.e., commits) are merged into the original project.

The provided dataset \cite{Gousi13} includes 108718 projects, 499485 users, 150362 issues, 78955 pull requests and 555325 commits among other numbers. It is important to note that only 91 of them are original projects (i.e., they are not a fork from another project). 

\begin{figure}[!t]
\centering
\includegraphics[width=\columnwidth]{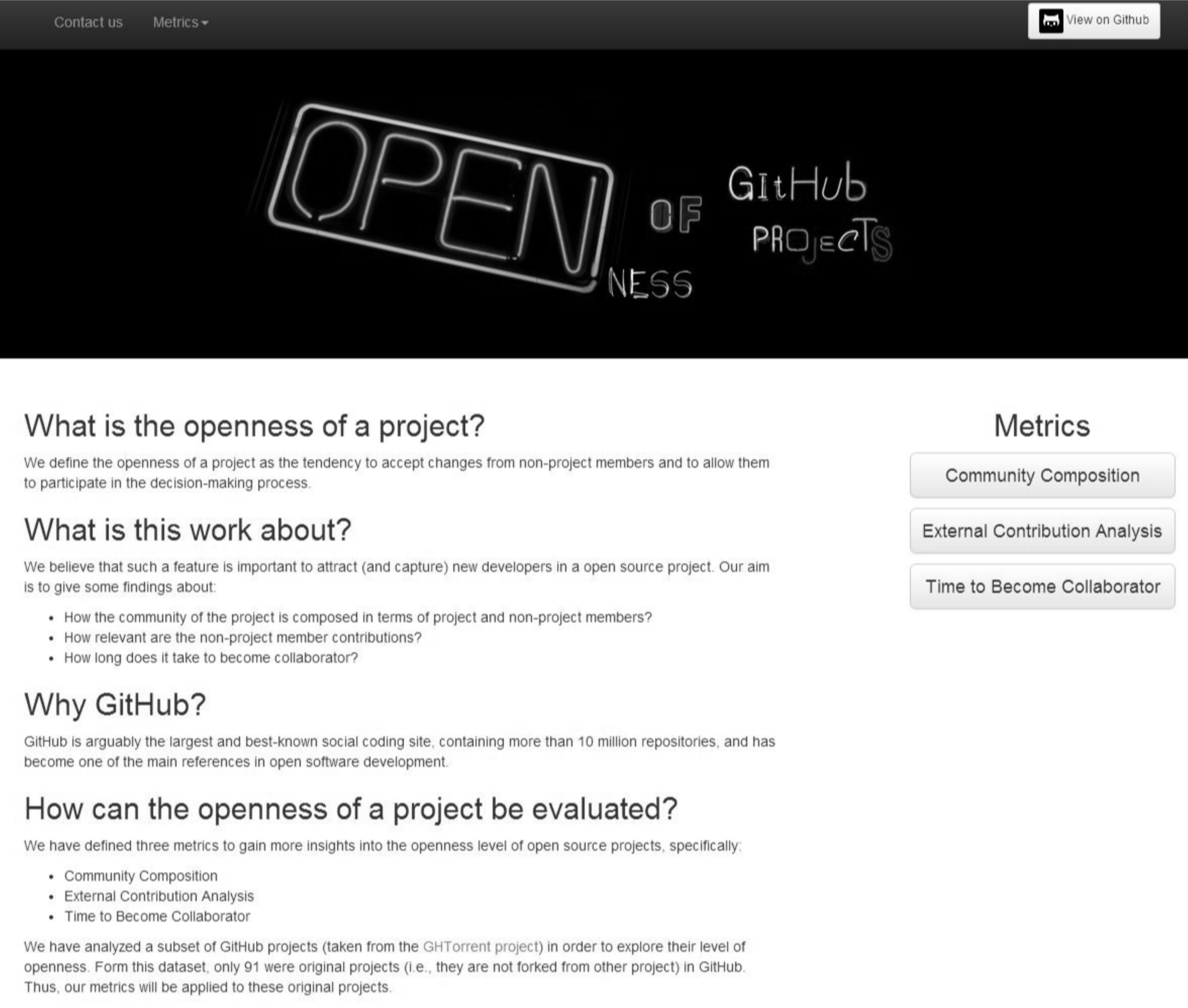}
\caption{Screenshot of the developed website including the metric results for the dataset. }
\label{fig:sc}
\end{figure}

Next sections will present the metrics and the results obtained. We will motivate briefly each metric, describe how it can be calculated from the data provided and finally report on the results obtained. The complete report including the results of the metrics for each original project can be found at \texttt{http://atlanmod.github.io/openness}. Figure \ref{fig:sc} shows a screenshot of the developed website.

\section{Community Composition}
\label{sec:op1}
People involved in the development of open source projects can play different roles. The analysis of the project community and its distribution will help us to understand the importance given to the participation of non-project members in the project, which is one of the openness factors in our view.

Based on the provided GitHub dataset we have identified four groups of users. Below, we describe each group and how to identify its members from the dataset: 
\begin{description}
  \item Project members (including the project owner), who are officially part of the project. They can be easily collected by querying the corresponding tables (i.e., \texttt{project\_mem} \texttt{bers} and \texttt{projects}) in the dataset.
  \item Collaborators, who are granted with the permission to manage the project, but are not part of the project members group. Collaborators are listed in the GitHub project webpage but they don't explicitly appear as such in the dataset. Therefore, collaborators are inferred from the dataset by looking for non-project members that have performed some management action (e.g., merging or closing a pull request, closing or reopening an issue, or performing an intra-branch pull request\footnote{The complete list of actions can be found at \texttt{https://help.github.com/articles/what-are-the-diffe} \texttt{rent-access-permissions}.}).
\item External contributors, who perform pull requests but do not have the permission to accept or close them. This group of users is not represented explicitly in the dataset but can be calculated by collecting those users who sent at least a pull request to an original project and are not a collaborator.  
\item External users, everyone else that contributes to the project (e.g., reporting or commenting an issue, etc.), but is not included in the previous groups.
\end{description} 

We call non-project members the union of collaborators, external contributors and external users. We differentiate between external contributors and external users to better study who is actually trying to contribute to the source code of the project.

\begin{figure}[!t]
\centering
\includegraphics[width=\columnwidth]{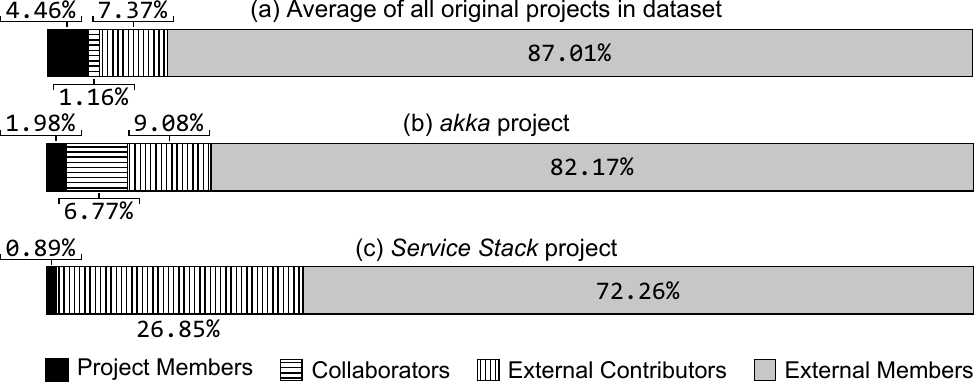}
\caption{Community composition. (a) shows the average values for the original projects in the dataset, (b) the results for the project \texttt{akka} and (c) for the project \texttt{ServiceStack}. }
\label{fig:op1}
\end{figure}

\vspace{1ex}
\noindent \textbf{Results}. We have applied the previous classification to every original project of the GitHub dataset and calculated the average values. The average result regarding the community composition for all the original projects is shown in Figure \ref{fig:op1}a. As can be seen, in average only 13\% of the community is willing to contribute to the code (the sum of project members, collaborators and external contributors). It is also important to note that the role of collaborator is scarcely used. This could be interpreted as that most projects prefer to keep a centralized authority (an exception would be the \texttt{akka} project also shown as example in Figure \ref{fig:op1}b). The number of external contributors could also be regarded as low but there also some important exceptions (like \texttt{ServiceStack}, in Figure \ref{fig:op1}c) showing that some projects do a good work in attracting volunteers. 

The analysis of these external contributors is the focus of the next metric.

\section{External Contribution Analysis}
\label{sec:op2}
A key openness indicator is the level of contributions coming from external contributors. To evaluate this metric, we define two different dimensions, the percentage of accepted external contributions and the time it takes to reach a decision (accept/reject) on a pull request \footnote{We only count for this metric pull requests coming from external contributors. Project members and collaborators can also send pull requests if they wish. These ``internal'' pull requests are filtered out in the computation of this metric}.

To calculate this metric we collect the pull requests from external contributors for each project. Then we calculate the elapsed time for all pull requests from the opening to the closing event, and finally we count how many of them were accepted/rejected. Note that pending pull requests are not considered. For each project we get the percentage of accepted/rejected pull requests and the average time to close them. Below, we present the results of this metric for the provided dataset.

\begin{figure}[!t]
\centering
\includegraphics[width=\columnwidth]{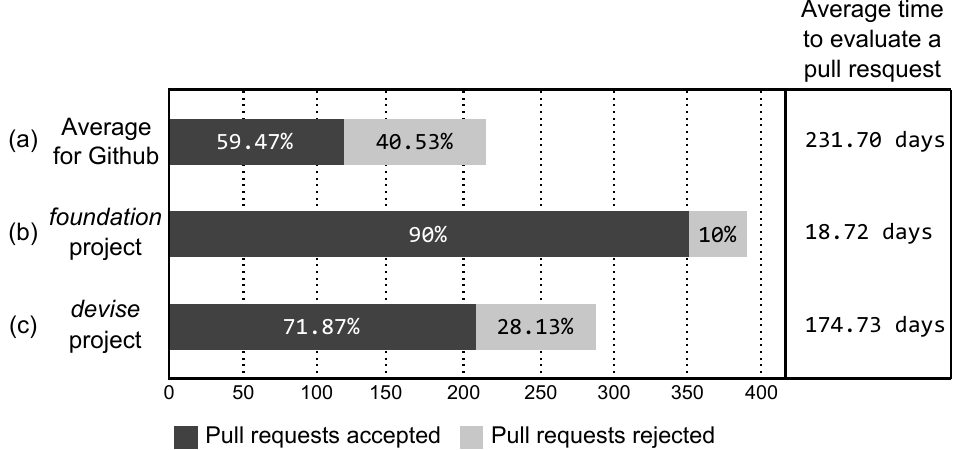}
\caption{Results of the external contribution analysis.  (a) shows the average values for the original projects in the dataset, (b) the results for the project \texttt{foundation} and (c) for the project \texttt{devise}.}
\label{fig:op2}
\end{figure}

\vspace{1ex}
\noindent \textbf{Results}. Figure \ref{fig:op2} shows the results obtained for this metric. On average, 59.47\% of pull requests are accepted and it takes around 231.70 days to address them. That it takes so long so evaluate a pull request came as a surprise and poses some questions about the ``agility'' of open source projects. Obviously, these numbers vary a lot from project to project. Thus, we have selected the projects \texttt{foundation} and \texttt{devise} as illustrative examples. The \texttt{foundation} project accepts almost all pull requests (90\%) and it does it really fast. This is even more impressive given the number of pull requests it gets (almost 400 so far). On the other hand, the \texttt{devise} project takes more time to deal with a lower amount of pull requests. 

\section{Time to Become Collaborator}
\label{sec:op3}
The long-term engagement of developers in open source projects is crucial for the project to be alive. One way to favour this commitment is by means of rewarding volunteers. In GitHub the most important reward could be to be added to the list of official collaborator of the project. The willingness of project members to grant this kind management permission to external contributors is in our view an important openness indicator. To evaluate this metric, we analyze the average time it takes for external users to become collaborators. 

Nevertheless, calculating on this metric on the GitHub projects in the dataset is tricky due to the fact that the dataset neither includes a explicit list of collaborators, as we have explained in Section \ref{sec:op1}, for a project nor the timestamp of when an external user becomes collaborator. To make things even worse, we don't even have a timestamp for the moment on which somebody joins the project as user.

Therefore, we have adapted our metric to the available data. The average time to become a collaborator is computed as the time between the first management action of a user and the first registered action \footnote{We assume all collaborators start as external users; otherwise we cannot calculate the elapsed time. The metric filters out collaborators that, according to the data, cannot be identified first as external users} of that same user in the project. For instance, a developer could start first creating a new issue describing a bug then submitting pull requests to fix that bug and at some point in time we realize that the user is now performing actions that require the management permissions of a collaborator. At that point, we can identify him/her as collaborator and calculate the elapsed time between this moment at his/her initial contribution.

\begin{figure}[!t]
\centering
\includegraphics[width=0.9\columnwidth]{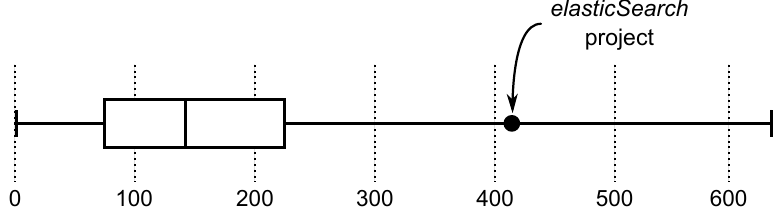}
\caption{Results for the collaborator group analysis. Median = 147.83. Q1 = 74.83. Q3 = 225.05. The result for the project \texttt{elasticSearch} (413.70) has been highlighted.}
\label{fig:op3}
\end{figure}

\vspace{1ex}
\noindent \textbf{Results}. The results of this metric are shown in the bloxplot of Figure \ref{fig:op3}. The average time is calculated only for those projects in which at least one external contributor became collaborator. The average results for the original projects in GitHub shows that the median value is 147.83 days. These values can be used to evaluate other projects in the dataset. For instance, the project \texttt{elasticSearch} has a value of 413.70, which is outside the box and may indicate a reluctance to grant management permission to external contributors.

\section{Related Work}
\label{sec:relatedWork}
Openness of software development projects has also been studied in other works such as \cite{Bein2010, Joode2006} but only from a ``static'' point of view (i.e. they analyze things like the language portability, library portability, license, etc.). Instead, we focus on openness from a community collaboration point of view.

The study of the community composition has also been explored in other works such as \cite{Gacek2004, OMahony2007}. In \cite{GPD14} authors study how a pull-based development model works, thus reporting on increased opportunities for community engagement and decreased time to incorporate contributions. Some of these ideas have been integrated in our openness evaluation. 

GitHub has been the target of other studies such as \cite{Thung2013}, which analyzes the impact of transparency in GitHub, and \cite{Bissyande2013}, which reports on the popularity, interoperability and impact of programming languages in several projects in GitHub. Some of their metrics could be incorporated to our approach to refine our results.

\section{Conclusion and Future Work}
\label{sec:conclusion}
Social coding sites like GitHub host milions of open source projects but only a small percentage are popular and manage to attract volunteer developers. We believe that the openness of a project plays a key role in attracting and engaging developers. 

Therefore, this paper proposes three metrics that help to evaluate the openness of an open source project. These metrics have been applied on a dataset of GitHub projects. Project members and owners can use these metrics to improve the attractiveness of their projects; while external contributors can rely on these metrics to identify those projects that better fit their development needs. 

As further work, we would like to perform a statistical analysis on the results of applying the metrics to all projects (possibly also in relation to the size of the project community) to see how they compare to each other and to try to infer some threshold values that help us to transform the metrics into recommendations (i.e. over which values a given project could be qualified as open). This should include also projects hosted in other social coding sites (e.g., BitBucket, SourceForge, etc.). We would also like to extend our study to consider other dimensions such as the activity and ``happinness'' of the project community. Works already carried out in other social websites such as Stack Overflow \cite{Grant2013, Sinha2013} can be useful here.

\balancecolumns

\bibliographystyle{abbrv}
\bibliography{msrChallenge}  

\end{document}